\documentclass{article}
\usepackage{spconf,amsmath,graphicx}
\usepackage{placeins}
\usepackage{float}
\usepackage{amsfonts}
\usepackage{algorithm2e}
\usepackage{comment}
\newcommand{\nosemic}{\SetEndCharOfAlgoLine{\relax}}

\RestyleAlgo{ruled}

\title{UNFUSED : \underline{UN}supervised \underline{F}inetuning \underline{U}sing \underline{SE}lf supervised \underline{D}istillation}
\name{Ashish Seth$^{2\star}$, Sreyan Ghosh$^{1\star}$, S. Umesh$^{2}$, Dinesh Manocha$^{1}$ \thanks{\hspace*{-1mm}$^{\star}$These authors contributed equally to this work}}

\address{
    $^1$University of Maryland, College Park, USA\\
  $^2$Speech Lab, Department of Electrical Engineering, IIT Madras, Chennai, India\\
  }
\begin{document}
%
\maketitle
\begin{abstract}
In this paper, we introduce \textbf{UnFuSeD}, a novel approach to leverage self-supervised learning and reduce the need for large amounts of labeled data for audio classification. Unlike prior works, which directly fine-tune a self-supervised pre-trained encoder on a target dataset, we use the encoder to generate pseudo-labels for \emph{unsupervised fine-tuning} before the actual fine-tuning step. We first train an encoder using a novel self-supervised learning algorithm (SSL) on an unlabeled audio dataset. Then, we use that encoder to generate pseudo-labels on our target task dataset via clustering the extracted representations. These pseudo-labels are then used to guide self-distillation on a randomly initialized model, which we call \emph{unsupervised fine-tuning}. Finally, the resultant encoder is fine-tuned on our target task dataset. Through UnFuSeD, we propose the first system that moves away from generic SSL paradigms in literature, which pre-train and fine-tune the same encoder, and presents a novel self-distillation-based system to leverage SSL pre-training for low-resource audio classification. In practice, UnFuSeD achieves state-of-the-art results on the LAPE Benchmark, significantly outperforming all our baselines. Additionally, UnFuSeD allows us to achieve this at a $\approx$ 40\% reduction in the number of parameters over the previous state-of-the-art system. We make all our codes publicly available\footnote{https://github.com/Sreyan88/LAPE}.
\end{abstract}
\section{Introduction}
\label{sec:intro}

Self-Supervised Learning (SSL) has proven to be one of the biggest success of the past decade enabling deep learning models to learn useful representations under low-resource labeled data settings. SSL has been adopted successfully in speech~\cite{baevski2020wav2vec}, vision~\cite{grill2020bootstrap,he2020momentum}, and text~\cite{devlin2018bert} outperforming all prior-art trained with only labeled data supervision on several benchmark datasets \cite{yang2021superb}. Though current SSL models pre-trained using Masked Acoustic Modeling (MAM) have been shown to generalize well over speech tasks like Automatic Speech Recognition (ASR), Phoneme Recognition (PR), etc., they fail to perform well on non-speech tasks like acoustic scene classification \cite{mesaros2018multidevice}. We list some possible reasons for this phenomenon in Section \ref{sec:related_work}. Thus, we emphasize the importance of learning general-purpose audio representations that can generalize over both speech and non-speech tasks, which is currently largely understudied in the literature compared to SSL in speech using MAM.

In the recent past, researchers have proposed novel algorithms for learning general-purpose audio representations \cite{9868132,saeed2021contrastive,niizumi2021byol}. A common trait among all these systems is that they directly fine-tune the model post-SSL pre-training. However, this direct fine-tuning approach may result in sub-optimal performance due to significant discrepancy between the pre-training and fine-tuning domains \cite{lee2022self}. For example, most of these systems perform SSL pre-training on the AudioSet \cite{gemmeke2017audio} (every day sounds like the sound of a toothbrush) and evaluate their learned representations on tasks like Speaker Verification \cite{Voxforge.org} (human spoken utterances). Additionally, under the linear evaluation setup, we argue that the downstream tasks cannot leverage the SSL representations to their full extent due to their learning capacity being constrained to an affine transform. 

\begin{figure*}[t!]
  \centering
  \includegraphics[width=0.8\textwidth]{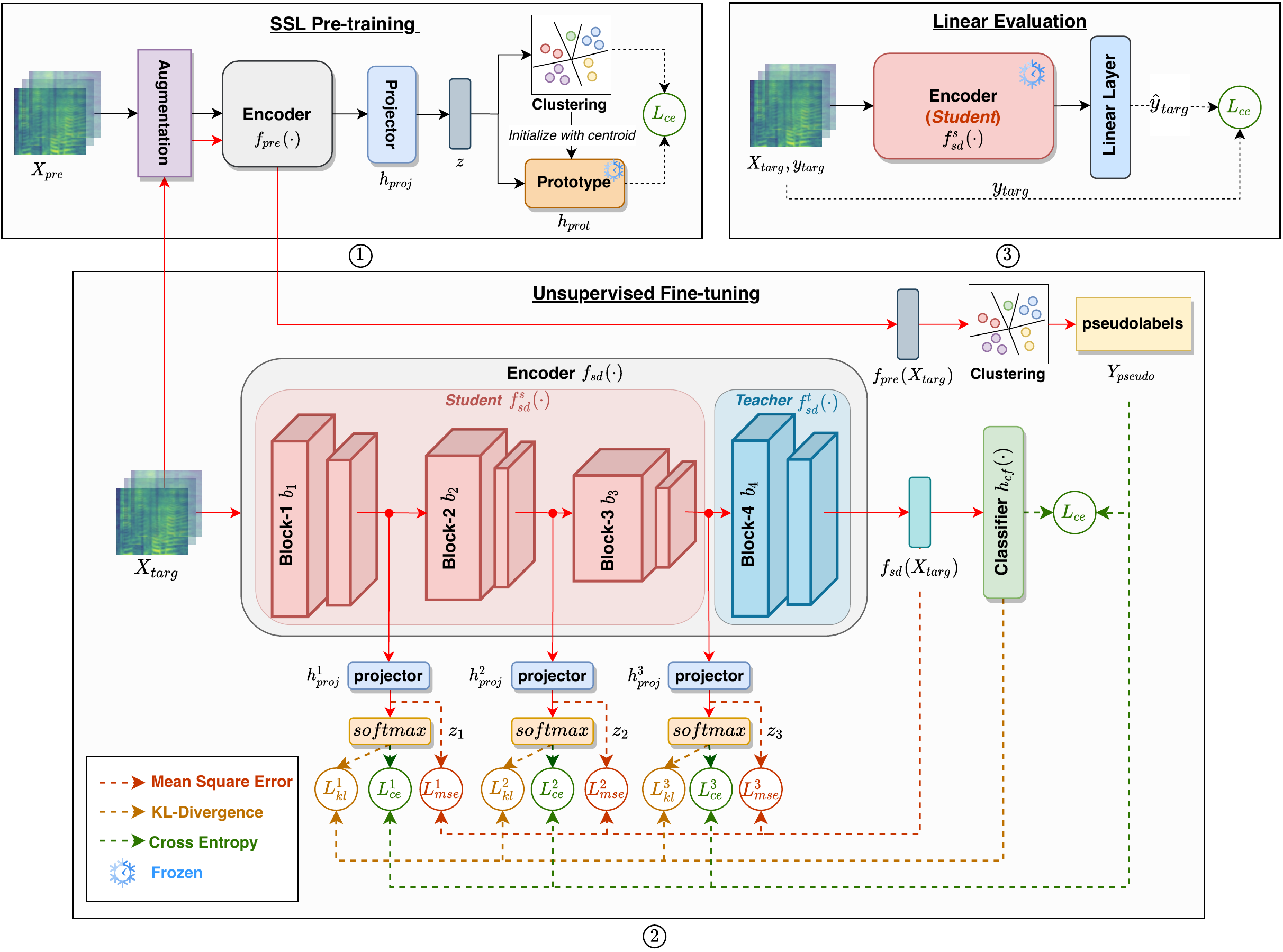}
  \caption{\small Illustration of \textbf{UnFuSeD}: UnFuSeD follows a 3 step training process from upstream SSL pre-training to downstream task-specific fine-tuning. \textcircled{\raisebox{-0.9pt}{1}} \textbf{SSL pre-training.} We first pre-train an convnet using un-labeled audio using DECAR-v2 (described in Section \ref{sec:method}).\textcircled{\raisebox{-0.9pt}{2}} \textbf{Unsupervised Fine-tuning.} We now pass the downstream task-specific data through the upstream model pre-trained in the last stage and extract and cluster these representations to generate pseudo-labels. These pseudo-labels are then used to perform \emph{unsupervised fine-tuning} on a randomly initialized convnet. \textcircled{\raisebox{-0.9pt}{3}} \textbf{Linear Evaluation.} Finally, a task-specific linear head is added to the convnet obtained from the previous step, and we perform supervised fine-tuning on the task-specific labeled dataset keeping the convnet frozen.}
  \label{fig:figure_1}
\end{figure*}

{\noindent {\bf Main Contributions:}} We present UnFuSeD, a new framework to improve downstream audio classification performance in low-resource labeled data settings leveraging SSL. Unlike all prior systems in literature, UnFuSeD does not directly fine-tune an SSL pre-trained model but uses it to extract and cluster audio features to generate pseudo-labels on a downstream task dataset which is then used to perform \emph{un-supervised fine-tuning}. More precisely, we perform a step of \emph{self distillation}, guided by the generated pseudo-labels, on a randomly initialized convnet encoder, divided into student and teacher encoders. Finally, post unsupervised fine-tuning, we perform supervised fine-tuning and evaluate downstream task performance on our model \emph{linear evaluation} setup. Additionally, to pre-train our encoder using SSL, we propose a novel SSL algorithm by modifying over DECAR \cite{ghosh2021deep}. Fig.\ref{fig:figure_1} shows a clear pictorial representation of our complete training process. We emphasize that UnFuSeD changes the paradigm in which SSL is leveraged to tackle data scarcity and improve downstream task performance. In practice, UnFuSeD achieves state-of-the-art (SOTA) performance on the LAPE Benchmark \cite{9868132} with an encoder with $\approx$ 40\% fewer parameters than the current SOTA model on LAPE.

\section{Related Work}
\label{sec:related_work}

{\noindent \textbf{Self Supervised Learning in Speech and Audio.}} The past decade has seen massive success in self-supervised learning in vision (CV), speech (SLP), and text (NLP), pushing the boundaries of low-resource representation learning for downstream classification \cite{baevski2020wav2vec,devlin2018bert}. The most common systems for SSL with speech solve a Masked Acoustic Modelling (MAM) task, either using contrastive learning \cite{baevski2020wav2vec}, frame reconstruction \cite{liu2020mockingjay}, or pseudo-label prediction \cite{hsu2021hubert}. However, recent research has shown that solving MAM makes the model representations mimic human articulatory responses \cite{wu2023speaker}, thus making it unsuitable for non-speech tasks. Thus, in the recent past, researchers have proposed novel systems to learn audio representations that can generalize over both speech and non-speech tasks. Following SSL in speech, these systems either solve a contrastive learning-based instance discrimination task \cite{saeed2021contrastive}, a clustering-based pseudo-label prediction task \cite{ghosh2021deep}, or a reconstruction task \cite{niizumi2021byol}. Knowledge distillation has shown great success in CV, and NLP with major applications in model compression \cite{10.1007/s11263-021-01453-z}. In a supervised setting, researchers have explored knowledge distillation for automatic speech recognition, speech emotion recognition, and speaker verification \cite{liu2022self}. DistillHubert \cite{chang2022distilhubert} was the first work on distilling SSL-based speech models and performs layer-wise knowledge-distillation (KD) to compress a full HuBERT \cite{hsu2021hubert}. On the other hand, when both encoder architectures are the same, this is known as self distillation (SD) \cite{pham2022revisiting}, and has shown impressive results with the student often outperforming the original teacher. However, to our knowledge, no existing work leverages SD for general-purpose self-supervised audio representation learning, and we are the first to explore this through UnFuSeD.


\section{Methodology}
\label{sec:method}

Fig.\ref{fig:figure_1} illustrates our proposed \textbf{UnFuSeD} learning algorithm. Algorithm \ref{alg:two} provides a detailed algorithmic overview of the same. In practice, UnFuSeD has three main steps, namely, (1) Upstream SSL Pre-training, (2) Unsupervised Fine-tuning, and (3) Downstream Supervised Fine-tuning. In the next paragraphs, we describe each step in detail.
\vspace{0.5mm}

{\noindent \textbf{(1) Upstream SSL Pre-training.}} Let $X_{pre}$ be an unlabeled dataset of size $J$ where $X_{pre}$ = $\{x_1, \cdots, x_j, \cdots ,x_J\}$ . In our case, here $J$ =  0.25 million, following the exact pre-training setup proposed by the LAPE benchmark. Our primary aim is to learn general-purpose audio representations from this unlabeled audio dataset. To achieve this, we use a simple convnet-based architecture \cite{koizumi2020ntt,takeuchi2020effects}, popular in prior-art \cite{9868132,niizumi2021byol} for a fair comparison. For upstream SSL pre-training, we propose DECAR-v2, an improved version of DECAR \cite{ghosh2021deep}, based on findings in \cite{caron2020unsupervised}. DECAR-v2 has two main steps or phases: (a) Assignment Phase and the (b) Training Phase.
\\
\textbf{(a) Assignment Phase:} The primary purpose of this phase is to obtain ``pseudo-labels'' $\mathbf{q}$ for every unlabelled audio sample $x \in X_{pre}$. To achieve this, we first store all the embeddings $\mathbf{g_{\tilde{x}}}$ obtained from our convnet projection head $h_{proj}$ in memory for the entire $X_{pre}$. After this, we apply Spherical K-means to cluster and get the ``pseudo-labels'' $\mathbf{q}$ for every $x$ as follows: $\min _{\mathbf{C} \in \mathbb{R}^{d \times K}} \frac{1}{N} \sum_{n=1}^{N} \min _{\mathbf{q}}-\mathbf{g_{\tilde{x}}}^{\top} \mathbf{C q}$ where $\textbf{C}$ is the Centroid matrix. Both $\mathbf{g_{\tilde{x}}}$ and columns of $\textbf{C}$ are $l_2$ normalized. $K$ represents the number of clusters, and $\tilde{x} \approx x$ is an augmented and sampled version of the original audio sample. Additionally, for ConvNet training stability, we keep the prototype head $h_{prot}$ parameters frozen throughout pre-training, and at the end of every assignment phase, the parameters of $h_{prot}$ are replaced by $\textbf{C}$.\\ 
\textbf{(b) Training Phase:} We train the network using supervision from the 
``pseudo-labels'' $\mathbf{q}$ obtained from the assignment phase. To do this, we first obtain the prediction $\mathbf{p}$ using $softmax(z)$ where $z$ is the output of the $h_{prot}$. Post this step; we minimize the multinomial logistic loss between $\mathbf{p}$ and $\mathbf{q}$ with: $\ell(\mathbf{p}, \mathbf{q})=-\sum_{k} \mathbf{q}^{(k)} \log \mathbf{p}^{(k)}$. The ``pseudo-labels'' are kept fixed during the training phase and updated for the entire $\textbf{X}$ only once every epoch during the assignment phase. Similar to \cite{caron2020unsupervised}, the assignment phase and training phase take place in isolation only at the first epoch, after which we use the embeddings $\mathbf{g_{\tilde{x}}}$ obtained from the previous epoch. These embeddings are stored in memory at every iteration of an epoch right after the back-propagation step.  
\vspace{0.5mm}

{\noindent \textbf{(2) Unsupervised Downstream Fine-tuning.}} After SSL pre-training, we don't fine-tune the pre-trained convnet $f_{pre}$ on the target task dataset directly but instead, use it for \emph{unsupervised fine-tuning} on a randomly initialized convnet $f_{sd}$. We call this step \emph{unsupervised fine-tuning} as we use the target task dataset but without using its actual labels. Let $D_{targ} = \{X_{targ}, Y_{targ}\}$ be the target task labeled dataset of size $I$ where $Y_{targ}$ are labels associated with audio samples $X_{targ}$. For unsupervised fine-tuning, we first generate $Y_{pseduo}$ by extracting and clustering representations obtained on passing $X_{targ}$ through $f_{pre}$. DECAR-v2 generates clusterable embeddings, which helps in the $Y_{pseudo}$ generation. We then use $Y_{pseduo}$ to perform self-distillation on $f_{sd}$. We first divide $f_{sd}$, which follows a similar architecture to $f_{pre}$, into a student ($f^{s}_{sd}$) and teacher network ($f^{t}_{sd}$). $f_{sd}$ has 4 individual blocks, where the first 3 make $f^{s}_{sd}$ and the last block makes $f^{t}_{sd}$. For more details on the architecture of $f^{s}_{sd}$, we refer our readers to \cite{9868132,koizumi2020ntt}. For self-distillation, we treat each block $b_i$ as a separate classifier and add a linear transform $h^i_{proj}$ to $b_i$ to solve three losses parallelly, KL-divergence $L_{kl}$, Mean-Square error $L_{mse}$ and Cross Entorpy $L_{ce}$. $L_{ce}$ ensures that the student blocks correctly classify the pseudo labels $Y_{pseduo}$ and thus utilize the weak supervision knowledge hidden in them. $L_{mse}$ ensures that knowledge of the deepest layers is leveraged to improve feature extraction in shallow layers. $L_{kl}$ ensures that the classification results of student classifiers are similar to that of the teacher classifier. Finally, to optimize our network, we use a weighted average of $L_{kl}$, $L_{mse}$ and $L_{ce}$, which we weigh by $\alpha$, $\beta$ as shown:
\vspace{-5.3mm}
\begin{equation}
L_{all}=L_{ce}+\alpha\sum^3_{i=1}L^i_{ce}+(1-\alpha)\sum^3_{i=1}L^i_{kl}+\beta\sum^3_{i=1}L^i_{mse}
\end{equation}
\vspace{-2.65mm}
\begin{equation*}
\begin{aligned}
    L_{ce} \leftarrow L_{ce}(l, Y_{pseudo}), L^i_{ce} \leftarrow L^i_{ce}(z^i, Y_{pseudo}),\\
    L^i_{kl} \leftarrow L^i_{kl}(z^i,l), L^i_{mse} \leftarrow L^i_{mse}(z^i, f_{sd}(X_{targ}))) \\where\, z^{i} = h^{i}_{proj}(f^{i}_{sd}(X_{targ})), l = h_{cl}(f_{sd}(X_{targ}))
\end{aligned}
\end{equation*}

\begin{algorithm}[t]
\linespread{0.85}
\caption{UnFuSeD}\label{alg:two}
\CommentSty{// SSL-pretraining}\\
\begin{small}\KwData{dataset $\mathcal{X}_{pre}$; number of clusters $K$; epoch $\mathcal{E}$; batch size $\mathcal{N}$}\end{small}
\begin{small}
\For{$epoch = 1$ $to$ $\mathcal{E}$}{
  \nosemic Sample a mini batch $\mathcal{X}^n_{pre}$ from $\mathcal{X}_{pre}$\;
  \nosemic Perform augmentations on $\mathcal{X}_{pre}$ to get $\tilde{\mathcal{X}}_{pre}$\;
  \nosemic Compute feature embedding obtained from encoder $f_{pre}(\tilde{\mathcal{X}}_{pre})$ and obtain $z = h_{proj}(f_{pre}(\tilde{\mathcal{X}}_{pre}))$\;
  \nosemic Initialize weights of $h_{prot}$ with centroid matrix $C$ obtained by $Kmeans(z)$.\;
  \nosemic Compute $Y_{pseudo}$ for $\mathcal{X}_{pre}$ using $C$\;
  \nosemic Minimize the cross-entropy $L_{ce}(Y^n_{pseudo}, \hat{Y}^n)$ where $\hat{Y}^n = softmax(h_{prot}(z^n))$\;
  \nosemic Update $f_{pre}$, $h_{proj}$ using gradient descent\;
}
\CommentSty{// Self-Distillation}\\
\KwData{target dataset $\mathcal{X}_{targ}$; number of classes $t$; epoch $\mathcal{E}$; batch size $\mathcal{N}$}
\For{$epoch = 1$ $to$ $\mathcal{E}$}{
\nosemic Sample a mini batch $\mathcal{X}^n_{targ}$ from $\mathcal{X}_{targ}$\;
\nosemic Compute $Y_{pseudo}$ using $Kmeans(f_{pre}(\mathcal{X}_{targ}))$ where $K=t$\;
\nosemic Compute $z^{i}$ = $h^{i}_{proj}(f^{i}_{sd}(\mathcal{X}^n_{targ}))$ and $l$ = $h_{cl}(f_{sd}(\mathcal{X}^n_{targ}))$
for each Block $b_i$ where $i \in \{1,2,3\}$\;
\nosemic Compute Cross-Entropy $L^i_{ce}(z^{i}, Y^n_{pseudo})$, KL-divergence $L^i_{kl}(z^i,l)$ and MSE $L^i_{mse}(z^i,f_{sd}(X_{targ}))$ loss for each Block $b_i$ where $i \in \{1,2,3\}$ (use Eq:1)\;
\nosemic Combine all losses $L_{all}$ with appropriate parameters $\alpha$, $\beta$ as stated in Eq:1\;
\nosemic Update $f_{sd}$, $h^{i}_{proj}$ for each Block $b_i$ where $i \in \{1,2,3\}$ and $h_{cl}$ using gradient descent\;
}
\end{small}
\end{algorithm}

\begin{table*}[t]
\centering
\caption{Result comparison of various SSL methods with proposed method \textbf{DECAR-v2} and \textbf{UnFuSeD} on the \emph{linear evaluation setup} with frozen encoder. The best results for each task are presented in bold. UnFuSeD outperforms all our baselines.}
\vspace{1mm}
\resizebox{\linewidth}{!}{%
\begin{tabular}{|l c c c c c c | c c|}
\hline 
\textbf{DT} & \textbf{BYOL-A} & \textbf{SimCLR} & \textbf{DECAR-v1} &  \textbf{DeLoRes-S} & \textbf{MoCo} & \textbf{DeLoRes-M} & \textbf{DECAR-v2} & \textbf{UnFuSeD}\\
\hline
\textit{Speech}& & & & & & & &\\
\hline

SC-V1 & $-$& 77.3 & 82.3 & 86.1 & $93.6 $ & $ 94.0$ & 91.6 &\textbf{94.4}\\

SC-V2(12) & 91.0 & 77.2 & 83.0 & 85.4 & $93.2 $ & $ 93.3$ & 90.6 &\textbf{94.1}\\

SC-V2(35) & 92.2 & 66.0 & 73.6 & 80.0 & $89.3 $ & $ 89.7$ & 87.2 & \textbf{90.1}\\

LBS &$-$& 89.0 & 91.0 & 90.0 & $95.5 $ & $ 95.7$ & 92.5 & \textbf{97.0}\\

VC & 40.1 & 28.9 & 25.6 & 31.2 & $42.5 $ & $ 45.3$ & 33.0 & \textbf{50.0}\\

IC &$-$& 59.8 & 63.2 & 60.7 & $65.1 $ & $ 65.2$ & 65.2 & \textbf{66.0}\\

VF & 90.2 & 69.2 & 74.1 & 76.5 & $87.3 $ & $ 88.0$ & 78.2 & \textbf{89.8}\\
\hline
\textit{Non-Speech} & & & & & & & &\\
\hline

NS & 74.1 & 61.3 & 70.7 & 66.3 & $74.7 $ & $75.0$ & 69.8 & \textbf{76.4}\\

BSD & $-$& 85.2 & 87.7 & 86.7 & $89.0 $ & $ 89.6$ & 88.5 & \textbf{90.0}\\

TUT &$-$& 52.4 & 62.5 & 58.6 & $66.7 $ & $ 65.7$ & 64.6 & \textbf{66.8}\\

US8K & 79.1 & 69.1 & 70.1 & 71.2 & $81.2 $ & $ 82.7$ & 73.2 & \textbf{83.2}\\
\hline
\textbf{Average} & $-$ & 66.9 & 71.2 &  72.1 & $79.8 $ & $ 80.4$ & 75.8 & \textbf{81.6}\\
\hline

\end{tabular}%
}
\label{table:without_barlow}
\end{table*}
{\noindent \textbf{(3) Supervised Downstream Fine-tuning}} Post unsupervised downstream fine-tuning, we do supervised downstream fine-tuning on the student model $f_{sd}$ using ${D}_{targ}$. For a fair comparison with prior-art in this space, we don't train all the layers of our model and instead just train a task-specific linear head added to the encoder. This method of training is also known as \emph{linear evaluation} and proves to be an effective technique for evaluating learned audio representations.

\section{Experimental Setup}
\label{sec:page}

{\noindent \textbf{Datasets.}} In our experiments, we use the exact same upstream and downstream training setups proposed by LAPE \cite{9868132}. For SSL-based pre-training, we use a balanced subset of 10\% of the complete AudioSet (0.2 million) and the FSD50K \cite{fonseca2021fsd50k}. For downstream tasks (DT), we evaluate our learned representations on LibriSpeech (LBS) \cite{7178964} and VoxCeleb (VC) \cite{Nagrani_2017} for speaker identification, Speech Commands (SC) v1 and v2 \cite{warden2018speech} for keyword spotting, VoxForge (VF) \cite{Voxforge.org} for language identification,  IEMOCAP (IC) \cite{busso2008iemocap} for speech emotion recognition, NSynth \cite{engel2017neural} for TUT Urban \cite{mesaros2018multidevice} and US8K \cite{10.1145/2647868.2655045} for acoustic event classification and finally Bird Song Detection (BSD) \cite{stowell2019automatic}.

{\noindent \textbf{Hyperparameter Tuning.}} For SSL Pre-training (DECAR-v2), we find the optimal values for the number of clusters as 512, learning rate as 0.005, batch size as 512, and number of epochs as 100. Projector $h_{proj}$ performs a $\mathbb{R}^{2048} \rightarrow \mathbb{R}^{512}$ non-linear transformation using multiple linear-layers. For Unsupervised Fine-tuning, we use the learning rate as 0.007, batch size as 512, number of epochs as 50, $\alpha$ as 0.7, and $\beta$ as 0.003. $h_{cf}$ performs a $\mathbb{R}^{2048} \rightarrow \mathbb{R}^{t}$ linear transformation, where $t$ is number of classes in target dataset. Projectors $h^1_{proj}$, $h^2_{proj}$ and $h^3_{proj}$ perform $\mathbb{R}^{2048} \rightarrow \mathbb{R}^{t}$, $\mathbb{R}^{1024} \rightarrow \mathbb{R}^{t}$ and $\mathbb{R}^{512} \rightarrow \mathbb{R}^{t}$ non-linear transformations respectively. Finally, for Linear Evaluation, we use the learning rate as 0.001, batch size as 32, and number of epochs as 50. All the hyperparameter choices were made based on an extensive grid search while considering the average performance across all the downstream tasks.      

\section{Results and Result Analysis}
\label{sec:results}

As clearly evident from Table \ref{table:without_barlow}, UnFuSeD outperforms all other approaches in literature by a significant margin. Results of BYOL-A were borrowed from their original papers. SimCLR was proposed as the pre-training approach in COLA \cite{wang2022towards} and was repeated on our convnet encoder using LAPE upstream dataset settings. We hypothesize that the gap in results from the original paper may be due to using a powerful encoder and 10 $\times$ more data from the AudioSet used in the paper. Measuring the effect of change in encoders is beyond the scope of this paper. Our proposed DECAR-v2 outperforms the already proposed DECAR-v1 by a margin of \textbf{4.6\%} (averaged across all tasks). Additionally, UnFuSeD outperforms DECAR-v2 by a margin of \textbf{5.8\%} (averaged across all tasks). Owing to space constraints, we provide results of UnFuSeD with different SSL training frameworks on our GitHub. Additionally, our final convnet encoder $f^{s}_{sd}$ used for downstream task evaluation has $\approx$ 40\% fewer parameters than DeLoRes-M \cite{9868132} (current SOTA system on the LAPE Benchmark).

\vspace{-1mm}
\section{Conclusion}
\label{sec:conclusion}

In this paper, we propose UnFuSeD, a novel methodology to leverage SSL for low-resource audio classification. In practice, UnFuSeD significantly outperforms all other approaches in literature on the LAPE audio evaluation benchmark. Additionally, we propose a new SSL algorithm called DECAR-v2 to learn general-purpose audio representations from unlabeled data.

\vfill\pagebreak

\bibliographystyle{IEEEbib}
\bibliography{strings,refs}

\end{document}